\documentclass[preprint,aps,ams,showkey,floatfix]{revtex4}
\usepackage{graphicx}
\usepackage{amssymb}

\begin{document}

\textbf{\Large Estimate solar contribution to the global surface
warming using the ACRIM  TSI satellite composite.}\\

Nicola Scafetta$^{1}$ and Bruce J. West$^{1,2}$\\
$^{1}$Physics Department, Duke University, Durham, NC 27708, USA.
$^{2}$Mathematical \& Information Science Directorate, US Army
Research Office, Research Triangle Park, NC 27709, USA.\\




\textbf{We study, by using a wavelet decomposition methodology, the
solar signature on global surface temperature data using the ACRIM
total solar irradiance satellite composite  by Willson and
Mordvinov. These data present  a +0.047\%/decade trend between
minima during solar cycles 21-23 (1980-2002).  We estimate that the
ACRIM upward trend  might have contributed $\sim$10-30\% of the
global surface
temperature warming over the period 1980-2002.}\\


\section{Introduction}
Among the potential  contributors to climate change,  solar forcing
is by far the most controversial. The Sun can influence  climate
through
 mechanisms that are not fully understood but which can be linked to  solar variations of luminosity,
magnetic field, UV radiation, solar flares and modulation of the
cosmic ray intensity [Pap and Fox, 2004; Lean, 2005]. In addition,
there is also controversy about solar data. Figure 1 shows two
similar but not identical satellite composites of  total solar
irradiance (TSI) that cover solar cycles 21-23 (1980-2002): the PMOD
due to Fr\"{o}hlich and Lean [1998] and the ACRIM due to Willson and
Mordvinov [2003], respectively.

PMOD  has been widely used in geophysical research. According to
this composite,  TSI has been almost stationary (-0.009\%/decade
trend of the 21-23 solar minima [Willson and Mordinov, 2003]) and by
adopting it, or the equivalent TSI proxy reconstruction by Lean
\textit{et al.} [1995], some researchers and the IPCC [Houghton
\textit{et al.} (2001); Hansen \textit{et al.} 2002] deduced that
the Sun has not contributed  to the observed
 global surface warming of the past decades. Consequently,  the global surface
warming of $\Delta T _{1980-2002}=0.40\pm0.04~ K$ from 1980 to 2002
shown in Figure 2 could only be induced, directly or indirectly,  by
anthropogenic added green house gas (GHG) climate forcing.

Contrariwise, ACRIM  presents a significant upward trend
(+0.047\%/decade trend of the minima) during solar cycles 21-23
(1980-2002) [Willson and Mordvinov, 2003]. The purpose of this
letter is to estimate the contribution of this upward trend to the
global surface warming from 1980 to 2002, which covers one Hale
solar cycle.

\section{Climate models and data analysis}

The ACRIM upward trend is evaluated by
 calculating the difference between the
 TSI average during solar cycle 21-22 (1980-1991) ($1365.95 \pm 0.08 ~W/m^2$) and
the TSI average during solar cycle 22-23 (1991-2002) ($1366.40 \pm
0.03 ~W/m^2$). We find this difference to be
\begin{equation}
\Delta I_{sun}= 0.45\pm 0.10~W/m^{2}~.  \label{uptrend2}
\end{equation}
The errors bars are calculated using multiple TSI averages by
considering that the period of a solar cycle  spans between 10 and
12 years and by keeping fixed the extremum at 1991. Note also that
the upward ACRIM modulation during solar cycles 21-23 can be
minimally interpreted as a 22-year square waveform modulation, which
recalls a Hale solar cycle, with amplitude $\Delta I_{sun}$.

There exist at least two ways to estimate the Sun's influence on
climate. The first method  relies on climate models, such as energy
balance models [Wigley, 1988; Stevens and North, 1996;  Foukal
\textit{et al.}, 2004] or general circulation models [Houghton J.T.,
\textit{et al.} (2001); Hansen \textit{et al.} 2002]. The climate
model approach is problematic because the sun-climate coupling
mechanisms  are not fully understood and, therefore, cannot be
confidently included in the computational models [Hoyt and Schatten,
1997; Hansen \textit{et al.}, 2002; Pap and Fox, 2004].

A second approach,  adopted, for example, by Douglass and Clader
[2002],  attempts to estimate the climate sensitivity to solar
variation by directly studying the signature of the solar cycles
within the temperature data. This is a phenomenological approach but
it has the advantage of evaluating the total effect of the
Sun-Climate coupling without requiring a detailed knowledge of the
underlying physical and chemical
 mechanisms. Herein we adopt this philosophy using a
 methodology that differs from the linear regression analysis
implemented by Douglass and Clader [2002], for reasons explained
later.

 The climate sensitivity $\lambda$ to a
 generic radiative forcing  $\Delta F$  is
 defined as $ \Delta T=\lambda \Delta F$, where
$\Delta T$ is the average temperature change induced by $\Delta F$.
The radiative forcing associated with a change of TSI, $\Delta I$,
is traditionally obtained by averaging $\Delta I$ over the entire
surface of the Earth  and allowing for a fraction (albedo $a\approx
0.3$) of $\Delta I$ to be reflected away: $\Delta
F_{sun}=\frac{1-a}{4}~\Delta I$.
 However,  the above definition is not  optimal if, as is commonly believed, the Sun
affects  climate by means of  direct or indirect mechanisms over and
above that of the direct TSI forcing. Because  solar phenomena
present cycles and general patterns that  \emph{mimic} TSI patterns,
we hypothesize that, to a first-order approximation, TSI can be used
as a geometrical proxy for the overall solar activity and its
effects on climate. Moreover, there might be a dependence of this
response on frequency [Wigley,  1988]. Thus, we introduce the
following model for the \emph{total} climate sensitivity to the
\emph{total} solar activity:
\begin{equation}
\Delta T_{sun}  = \int_0^\infty Z(\omega)~\frac{d \Delta I}{d\omega}
~d \omega ~. \label{solarcond}
\end{equation}
The frequency-dependent function  $Z(\omega)$ is herein defined as
the \emph{total climate sensitivity} to solar variations. Note that
Douglass and Clader [2002] adopted a model in which the function
$Z(\omega)$ is a constant $k$ at all frequencies such that: $\Delta
T_{sun}=k~\Delta I$.

Douglass and Clader [2002] evaluated the climate sensitivity to
solar variation, $k=0.11\pm0.02~ K/(Wm^{-2})$, by using the PMOD TSI
composite and by means of a multiple linear regression analysis
based on a predictor for the temperature $T(t)$ of the form $
C(t)=f(t)+k_{1}I(t-\tau _{1})+k_{2}S(t-\tau _{2})+k_{3}V(t-\tau
_{3})$, where $t$ is the time, $f(t)$ is a linear function,
 $I(t-\tau _{1})$ is the
solar irradiance, $S(t-\tau _{2})$ is a measure of the El Ni\~{n}o
Southern Oscillation (ENSO) indexed by the SST anomalies, $V(t-\tau
_{3})$ is a measure of the volcano-aerosol signal, $\tau _{i}$ are
fixed lag-times that give the highest correlation between each
signal and the data, and the $k_{i}$ are the corresponding forcing
constants. However, the multiple linear regression analysis is not
optimal because  the parameters $k_{i}$ and $\tau _{i}$ might be
time-dependent and, in such a case, keeping them constant would
yield serious systematic errors in the evaluation of the parameters
$k_{i}$. Moreover, climate models predict that the climate
sensitivity to cyclical forcing increases at lower frequencies
because of the strong frequency-dependent damping effect of ocean
thermal inertia [Wigley,  1988; Foukal \textit{et al.}, 2004]. Thus,
Douglass and Clader [2002] evaluated the climate sensitivity to the
11-year solar cycle, but as we have discussed above,  the upward
ACRIM modulation during solar cycles 21-23 can be minimally
interpreted as a 22-year cycle modulation with amplitude given by
Eq. (\ref{uptrend2}). Therefore, we have to evaluate the climate
sensitivity to a 22-year cycle and then we can approximate Eq.
(\ref{solarcond}) as
\begin{equation}
\Delta T_{sun}  \gtrapprox  Z_{22years}~ \Delta I_{sun} ~.
\label{solarcond2}
\end{equation}

 We  proceed by decomposing the solar and temperature signals with proper
band-pass filters for isolating the frequency bands  of interest.
The purpose is to estimate a linear transfer coefficient $
Z(\omega)=A_{out}(\omega)/A_{in}(\omega)$ by comparing the amplitude
$A_{in}(\omega)$ of an oscillating input signal at a given frequency
$\omega$, with the amplitude $A_{out}(\omega)$ of the oscillating
output signal at the same frequency and then to apply Eq.
(\ref{solarcond2}). Linear transfer analysis is the usual method
adopted  to estimate the sensitivity of a complex but unknown system
to external stimulation.

The band pass filter we adopt is based on the maximal overlap
discrete wavelet transform (MODWT) multiresolution analysis (MRA) by
means of the 8-tap Daubechies least asymmetric (LA8) filter
[Percival and Walden, 2000]. MRA makes use of scaled waveforms that
measure signal variations by simultaneously analyzing the signal's
time and scaling properties and, therefore, can powerfully identify
local non-periodic patterns and signal singularities, and
characterize signal structures [Percival and Walden, 2000].  Thus,
the wavelet filtering is more efficient than the traditional linear
transport frequency filters for extracting patterns in the data.

MODWT MRA decomposes a time series $X(t)$ into a  hierarchical
sequence of zero-centered  band-pass filter curves called
\emph{detail curves} $D_{j}(t)$,  and a hierarchical sequence
of smooth low-pass filter curves, called $%
S_{j}(t)$. High-pass filter curves are referred to as residual
curves and indicated with $R_j(t)$. The index $j$ indicates the
order of scaling. So, at the $J^{th}$ order  MODWT MRA decomposes a
signal $X(t)$ as
$X(t)=S_{J}(t)+\sum_{j=1}^{J}D_{j}(t)=S_{J}(t)+R_{J}(t).$
 The smooth
curve $S_J(t)$ captures the smooth modulation of the data with a
time scale larger than $2^{J+1}$ units of the time interval $\Delta
t$ at which the data are sampled. The detail curve $D_{j}(t)$
captures  local variations with period approximately ranging from
$2^{j}\Delta t$ to $2^{j+1}\Delta t$. Finally, the residual curve
$R_J(t)=X(t)-S_J(t)=\sum_{j=1}^{J}D_{j}(t)$ captures  local
variations of the data at  time scales shorter than $2^{J+1}\Delta
t$.

The global surface temperature data are sampled monthly.  The
11-year cycle (132 months) would be captured by the wavelet detail
$D_{7}(t)$ that corresponds to the band between $2^7=128$ and
$2^8=256$ months. However, the solar cycles are pseudo-periodic and
to avoid an excessive random split of the cycles between adjacent
wavelet detail curves, the wavelet filter should be optimized by
choosing a time interval $\Delta t$ such that the 11-year
periodicity falls in the middle of the band captured by the curve
$D_{7}(t)$. The average between 128 and 256 is 192, and the correct
time interval is $\Delta t=132/192=0.6875$ months. By using a linear
interpolation  we transform the monthly  temperature data into a new
time series sampled at $\Delta t=0.6875$ months,
and then  apply the MRA to it. Thus, the detail curve $%
D_{7}(t)$ captures the scaling band between 88-176 months (or
7.3-14.7 years)  centered in the  11-year solar cycle, while the
detail curve $D_{8}(t)$ captures the band between 176-352 months (or
14.7-29.3 years)  centered in the  22-year solar cycle. Figure 3
shows the MODWT MRA of the global mean surface temperature since
1856 defined by the decomposition
\begin{equation}
T(t)=S_{8}(t)+D_{8}(t)+D_{7}(t)+R_{6}(t)~. \label{newfig3}
\end{equation}
The smooth curve $S_8(t)$ captures the secular variation of the
temperature at time scale larger than 29.3 years that is reasonably
produced by the slow modulation of the GHG and aerosol forcings plus
the slow secular variation of the solar forcing. The detail curves
$D_{8}(t)$ and $D_{7}(t)$ correspond, according to our hypothesis,
to the climate signature imprinted by the 22-year and 11-year solar
cycles respectively. The residual curve $R_{6}(t)$ collects all
climate fluctuations at a time scale shorter than 7.3 years, which
is mostly affected by SST oscillations, volcano eruptions and
undetermined noise.

Figure 4 compares the band-pass curves $D_{7}(t)$ and $D_{8}(t)$ for
the TSI data and global temperature anomalies. For the period
1856-1980 we apply the MRA to the TSI proxy reconstruction by Lean
\textit{et al.} [1995], while for the period 1980-2002 the MRA is
applied to  the ACRIM TSI. Several 11-year solar cycles are easily
recognizable in the corresponding $D_{7}(t)$ temperature cycles in
particular after 1960. The slow 22-year solar cycles seem even
better reproduced in the temperature detail curve $D_{8}(t)$ and the
temperature response lags the Hale solar cycles since 1900 by
approximately $2.2 \pm 2$ years.

We evaluate the linear transfer coefficient $Z_{7}$ and $Z_{8}$ by
estimating the amplitude of the solar and temperature oscillations
associated with the band-pass curves $D_{7}(t)$ and $D_{8}(t)$
during the period 1980-2002. The amplitude $A$ of an oscillating
signal, $f(t)=\frac{1}{2}A\sin (2\pi t)$, is related to the signal
variance $ \sigma^2 =\frac{1}{T}\int_{0}^{T}\left[
f(t)-\overline{f(t)}\right] ^{2}dt,$ where $T$ is the time period
and $\overline{f(t)}$ is the average of the signal, via the relation
$ A=2\sqrt{2}~\sigma$.

For the ACRIM data we find $ A_{7,sun}=0.92\pm 0.05~W/m^{2} $ and $
A_{8,sun}=0.35\pm 0.10~W/m^{2} $. For the temperature data we find
 (11-year signature) $A_{7,temp}=0.10\pm 0.01 ~K$ and (22-year signature) $%
A_{8,temp}=0.06\pm 0.01 ~K$. Thus, we obtain:
\begin{eqnarray}
  Z_{7}=A_{7,temp}/A_{7,sun} &=& 0.11\pm 0.02~ K/(Wm^{-2}), \label{ab7} \\
  Z_{8}=A_{8,temp}/A_{8,sun}&=& 0.17\pm 0.06~ K/(Wm^{-2}). \label{a788987}
\end{eqnarray}
Eqs. (\ref{ab7}) and (\ref{a788987}) refer to the climate
sensitivity to the 11-year and 22-year solar cycles from 1980 to
2002  using the ACRIM TSI composite, respectively.

\section{Discussion and conclusion}

Our methodology filtered off
 volcano-aerosol and ENSO-SST  signals
from the temperature data because these estimates are partially
consistent with already published independent empirical findings. In
fact, the 11-year climate sensitivity $Z_{7}=0.11\pm 0.02~
K/(Wm^{-2})$  is equal to the 11-year climate sensitivity $k$
estimated by Douglass and Clader [2002]. Douglass and Clader also
estimated that the 11-year solar cycle is associated with a $0.10~K$
temperature cycle and  this value is equal to our estimate
$A_{7,temp}$, see also Lean [2005]. Because Douglass and Clader used
a multiple linear regression analysis to separate the 11-year solar
signature from the volcano-aerosol and ENSO-SST signals we can
conclude that our wavelet band-pass filter has efficiently filtered
off from the temperature data both volcano-aerosol and ENSO-SST
signals. Evidently, from 1980 to 2002 volcano-aerosol and ENSO-SST
signals affected climate on time scales shorter than 7.3 years which
are captured by the residual curve $R_6(t)$.

Our climate sensitivities $Z_7$ and $Z_8$ were also approximately
anticipated by White \textit{et al.} [1997]. These authors, by
adopting Fourier band-pass filters centered at 11 and 22 year
periodicities  respectively, studied the response of global upper
ocean temperature to changing solar irradiance using the TSI proxy
reconstruction by Lean \textit{et al.} [1995] from 1900 to 1991.
Their regression coefficients between solar and temperature cycles
are $k_{11-years}=0.10\pm 0.02~ K/(Wm^{-2})$ and
$k_{22-years}=0.14\pm 0.02~ K/(Wm^{-2})$. These estimates are
slightly smaller than $Z_7$ and $Z_8$, respectively, probably
because these authors analyzed a different  temporal period, and
adopted a hypothetical TSI sequence and ocean surface temperature
while we used global surface temperature, and over land the climate
response to  solar variation is stronger than over ocean.

The climate sensitivity to the 22-year cycle, $Z_8$, is
approximately 1.5 times stronger than the climate sensitivity to the
11-year cycle, $Z_7$, and, in average, the 22-year climate response
lags Hale solar cycles by approximately $2.2\pm 2$ years. Both
effects are approximately predicted by theoretical energy balance
models. In fact, the actual climate response to cyclical forcing is
stronger at lower frequencies because the damping effect of the
ocean inertia is weaker at lower frequencies [Wigley 1988, table 1].
This frequency dependence arises because the system is typically not
in thermodynamic equilibrium. The ratio $ Z_{8}/Z_{7}=1.55\pm0.55 $
is consistent with that between the damping factors for 20 and 10
year periodicities $\eta_{20}/\eta_{10}\approx 1.45$ indicated by
the Wigley's model [1988, table 1]. Wigley's model  also predicts a
response-lag of 2.5-2.8 years for a 20 year periodicity.

In conclusion, we believe our estimates $Z_7$ and $Z_8$ of the
climate sensitivity to solar variations from 1980 to 2002 are
realistic. By using the ACRIM TSI increase estimate $\Delta I_{sun}$
(\ref{uptrend2}), the climate sensitivity $Z_8$ in Eq.
(\ref{a788987}) and Eq. (\ref{solarcond2}), the warming caused by
$\Delta I_{sun}$ is $ \Delta T_{sun} \gtrapprox 0.08\pm0.03$. Thus,
because the global surface warming during the period 1980-2002 was
$\Delta T _{1980-2002}=0.40\pm0.04~ K$, we conclude that according
to the ACRIM TSI composite the Sun may have minimally contributed
$\sim$10-30\% of the 1980-2002 global surface warming.

Lastly,  we compare the observed 11-year temperature cycle
amplitude, $A_{7,temp}=0.10\pm 0.01 ~K$, with that estimated by some
theoretical climate models. By adopting three energy balance models,
Stevens and North [1996] show in their figure 15 that 11-year TSI
cycle forcing since 1980 would imprint 11-year global surface
temperature cycles with an amplitude $A_{temp}\approx0.06\pm0.01K$;
the MAGICC climate model by Wigley gives $A_{temp}\approx 0.035K$
[Foukal \textit{et al.}, 2004]. Consequently, our estimate of the
11-year temperature cycle $A_{7,temp}$ is approximately 1.5-3 times
larger than what these models predict.  Douglass and Clader [2002]
arrived to a similar conclusion about the Wigley's model. Thus,
while the theoretical models approximately predict the relative
climate sensitivity ratio $Z_8/Z_7$ and the response time-lag, they
seem to disagree from each other about the actual climate
sensitivity to solar variation and significantly underestimate the
phenomenological climate sensitivities to solar cycles as we have
estimated. Evidently, either the empirical evidence deriving from
the deconstruction of the surface temperature is deceptive, or the
 models are inadequate because of the difficulty of
modeling climate in general and a lack of knowledge of climate
sensitivity to solar variations in particular. As Lean [2005] noted,
the models might be inadequate: (1) in their parameterizations of
climate feedbacks and atmosphere-ocean coupling; (2) in their
neglect of indirect response by the stratosphere and of possible
additional climate effects linked to solar magnetic field, UV
radiation, solar flares and cosmic ray intensity modulations; (3)
there might be other possible natural amplification mechanisms
deriving from internal modes of climate variability which are not
included in the models. All the above mechanisms would  be
automatically considered and indirectly included  in our
phenomenological approach.

\newpage

\begin{figure}
\includegraphics[angle=-90,width=20pc]{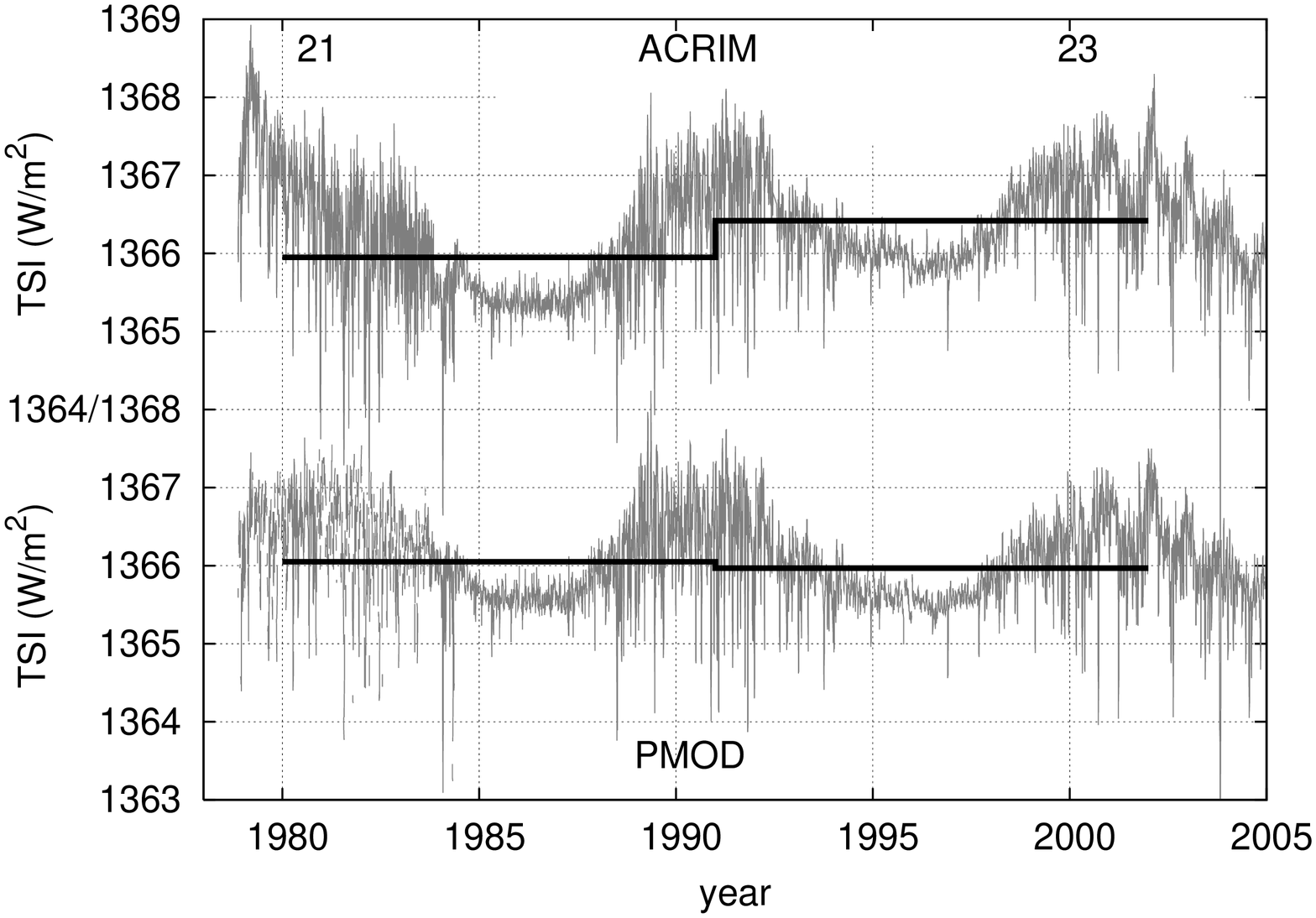}
\caption{  ACRIM  TSI composite  by Willson and Mordvinov [2003] and
an update of the PMOD TSI composite by Fr\"{o}hlich and Lean [1998].
The black lines are the TSI averages in the periods 1980-1991 and
1991-2002. }
\end{figure}

\begin{figure}
\includegraphics[angle=-90,width=20pc]{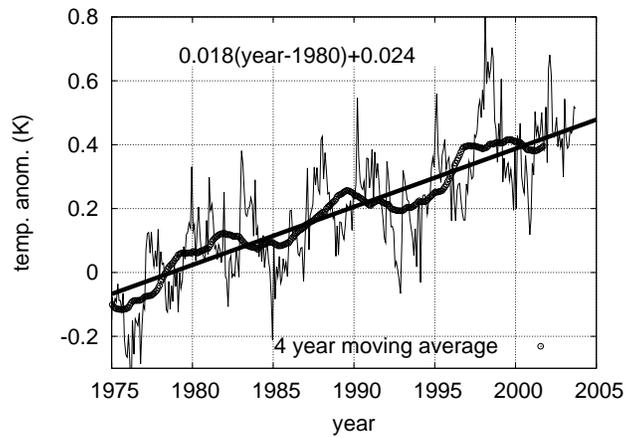}
\caption{ Global mean surface temperature anomalies.   The global
surface warming from 1980 to 2002, estimated with a linear fit, is
$\Delta T _{1980-2002}=0.40\pm0.04~ K $. Data from CRU (2005).}
\end{figure}

\begin{figure}
\includegraphics[angle=-90,width=20pc]{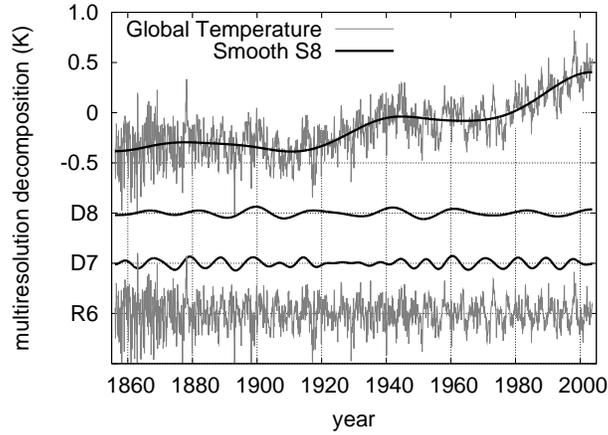}
\caption{ Global surface temperature (1856-2002)  [CRU, 2003] and
its MODWT MRA
 according to Eq. (\ref{newfig3}). The residual
curve $R_6(t)$ becomes progressively  less noisy probably because of
improved observations during the last 150 years. }
\end{figure}

\begin{figure}
\includegraphics[angle=0,width=20pc]{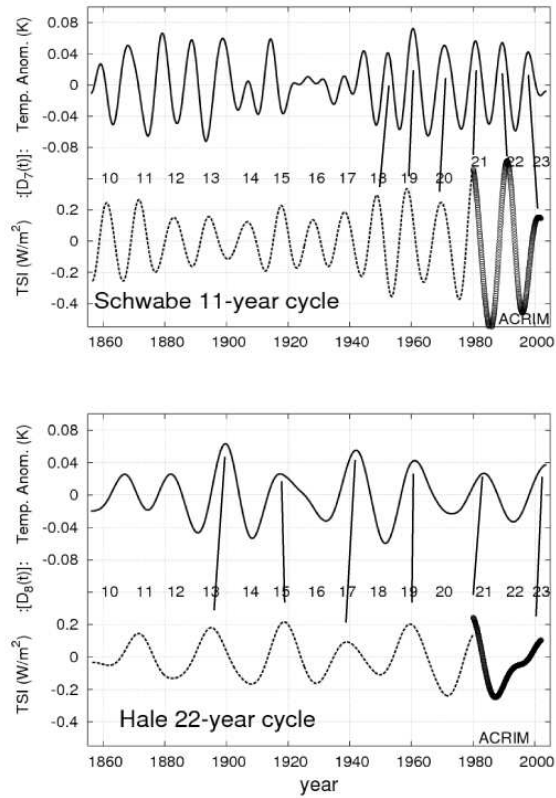}
\caption{ MODWT MRA band-pass curves $D_7(t)$ and $D_8(t)$ of global
temperature
(solid line) and TSI proxy reconstruction (1856-1980) by Lean \textit{et al.} [1995] (dash line). The \textit{%
`circle'} curve refers to the MODWT MRA band-pass curves applied to
the ACRIM TSI (1980-2002).}
\end{figure}


\end{document}